\documentclass[aps,prl,twocolumn,superscriptaddress,amssymb,amsmath,showpacs]{revtex4}
\usepackage{dcolumn}
\usepackage{graphicx}

\begin{document}


\title{Quasiparticle Effects in the Bulk and Surface-State Bands of Bi$\mbox{\boldmath $_2$}$Se$\mbox{\boldmath $_3$}$ and 
Bi$\mbox{\boldmath $_2$}$Te$\mbox{\boldmath $_3$}$ Topological Insulators}

\author{Oleg V. Yazyev}
\affiliation{Department of Physics, University of California, Berkeley, California 94720, USA}
\affiliation{Materials Sciences Division, Lawrence Berkeley National Laboratory, Berkeley, California 94720, USA}
\author{Emmanouil Kioupakis}
\affiliation{Materials Department, University of California, Santa Barbara, California 93106, USA}
\author{Joel E. Moore}
\affiliation{Department of Physics, University of California, Berkeley, California 94720, USA}
\affiliation{Materials Sciences Division, Lawrence Berkeley National Laboratory, Berkeley, California 94720, USA}
\author{Steven G. Louie}
\affiliation{Department of Physics, University of California, Berkeley, California 94720, USA}
\affiliation{Materials Sciences Division, Lawrence Berkeley National Laboratory, Berkeley, California 94720, USA}

\date{\today}


\begin{abstract}
We investigate the bulk band structures and the surface states of Bi$_2$Se$_3$ 
and Bi$_2$Te$_3$ topological insulators using first-principles many-body perturbation 
theory based on the $GW$ approximation. The quasiparticle self-energy corrections 
introduce significant changes to the bulk band structures, while their effect on the 
band gaps is opposite in the band-inversion regime compared to the usual situation without 
band inversion. Parametrized ``scissors operators'' derived from the bulk 
studies are then used to investigate the electronic structure of slab models which exhibit
topologically protected surface states. The results including self-energy corrections 
reveal significant shifts of the Dirac points relative to the bulk bands and large gap 
openings resulting from the interactions between the surface states across the thin slab, 
both in agreement with experimental data. 
\end{abstract}

\maketitle


The recently discovered bulk topological insulators (TIs), a new class of semiconducting 
materials characterized by the presence of spin-helical surface states resulting from strong
spin-orbit (SO) interactions, have quickly become a subject of intense research \cite{reviews}.
It is believed that TIs are actually not uncommon among the heavy-element materials. Owing to 
their novel electronic properties, such as the suppression of backscattering and intrinsic spin 
polarization of the surface-state charge carriers, TIs are expected to find applications in the 
future, including information technology, spintronics and quantum computing. 


First-principles electronic structure calculations can play an important role
in exploring the properties of known TIs as well as in guiding the search of novel materials.
The widely used density functional theory (DFT) within the Kohn-Sham formalism, a workhorse first-principles 
method of condensed matter physics, has proved its value already in the discovery of 
the ``second-generation'' bulk TIs, Bi$_2$Se$_3$ and Bi$_2$Te$_3$ \cite{Xia09,Zhang09}. A large 
number of TIs has been predicted using this technique \cite{TI_predictions}, and some of these 
predictions have been confirmed experimentally \cite{BiTlX2}.   
However, it is broadly recognized that the Kohn-Sham eigenvalues of DFT fail to describe accurately quasiparticle energies 
and band gaps \cite{Louie06}, the critical properties of TIs. A recent work has highlighted 
the limitations of standard DFT approach in describing the topological nature of several borderline compounds \cite{Vidal11}.
Many-body perturbation theory techniques, such as the $GW$ approximation, greatly improve the accuracy of predicting these 
excited-state properties \cite{Hedin70,Hybertsen86a}.  
   
In this Letter, we investigate the self-energy effects in the quasiparticle bulk band structures 
and the surface-state dispersion of the reference TIs, Bi$_2$Se$_3$ and Bi$_2$Te$_3$,
by using the first-principles $GW$ method \cite{Hybertsen86a}. We find that the effects of quasiparticle corrections on 
the band structures are substantial, and show novel features resulting from the 
interplay with SO interactions in the band-inversion regime. By introducing parametrized energy-dependent 
``scissors operators'' based on the bulk calculations, we have extended our study to the electronic structure of slab models.
Our application of the proposed technique to slab models which exhibit topologically protected
surface states finds significant shifts of the Dirac point energies relative to the bulk bands
and larger surface-state gap openings resulting from the interactions between surface states
across the slab. The proposed simple corrections basically eliminate the DFT eigenvalue problems and yield agreement with experimental
data.


The DFT calculations were performed within the local density approximation (LDA)
employing the {\sc Quantum-ESPRESSO} package \cite{QE}. We used norm-conserving 
pseudopotentials \cite{Troullier91} and a plane-wave kinetic energy cutoff of 35 Ry for the 
wavefunctions. The quasiparticle energies were evaluated within the $G_0W_0$ 
approximation to the electron self-energy starting from LDA results as a mean-field 
solution using the approach of Hybertsen and Louie \cite{Hybertsen86a}. The static dielectric function was calculated using a 10~Ry 
plane-wave cutoff, unoccupied bands up to 5~Ry above the Fermi level, and 
extended to finite frequencies with the generalized plasmon-pole model. This
first-principles $GW$ methodology is implemented in the {\sc BerkeleyGW} code \cite{BerkeleyGW}. 
Spin-orbit interactions were included on the final stage using the SO Hamiltonian 
matrix ${\mathcal H}_{\rm SO}({\bf k})$ evaluated employing an approach described in 
Ref.~\onlinecite{Hybertsen86b} in the basis of eigenfunctions of ${\mathcal H}_0 ({\bf k})$.
That is, in the full two-component Hamiltonian
\begin{equation}
\label{so}
{\mathcal H}({\bf k})={\mathcal H}_0 ({\bf k}) + {\mathcal H}_{\rm SO}({\bf k})
\end{equation}
${\mathcal H}_0 ({\bf k})$ is a diagonal matrix with matrix elements being either LDA or $GW$ 
eigenvalues in the absence of SO interactions, while ${\mathcal H}_{\rm SO}({\bf k})$
introduces off-diagonal matrix elements.
Notably, we find that the results obtained using this method are in agreement with an explicit approach in 
which $GW$ calculations are performed starting from two-component LDA wavefunctions 
after SO interactions were taken into account (and thus, band inversion is present 
at the $\Gamma$ point) \cite{Yazyev11}. We used experimental lattice parameters for both 
the bulk crystal as well as (111) slab models of different thickness. 
 
\begin{figure}
\includegraphics[width=7.25cm]{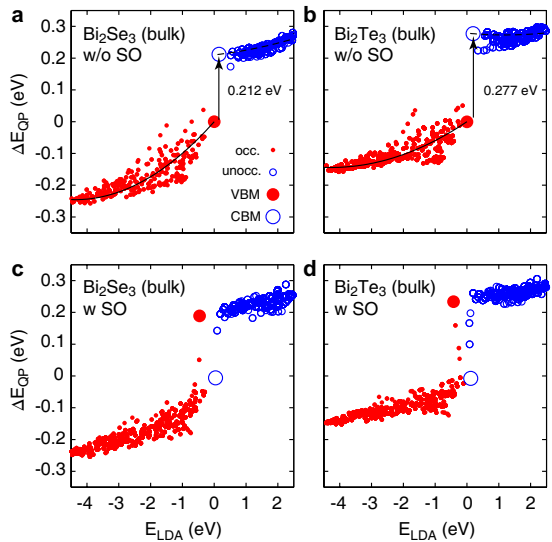}
\caption{\label{fig1}
(color online). Quasiparticle self-energy corrections as a function of 
LDA energies for bulk (a) Bi$_2$Se$_3$ and (b) Bi$_2$Te$_3$ calculated
without taking into account SO interactions. The positions of VBM and CBM
are indicated by the larger filled and open circles, respectively. The lines correspond to the fitted ``scissors operators''.
(c,d) Same plots after the SO matrix elements were taken into account.
Note the changes in $GW$ corrections calculated for the states which 
correspond to VBM and CBM in (a) and (b).
}
\end{figure}

\begin{figure}[b]
\includegraphics[width=7.5cm]{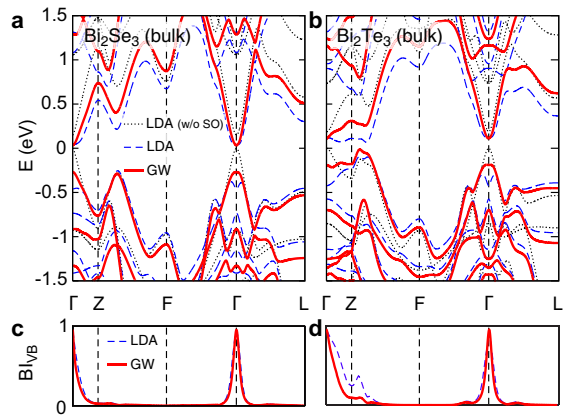}
\caption{\label{fig2}
(color online). Band structures of bulk (a) Bi$_2$Se$_3$ and (b) Bi$_2$Te$_3$ calculated
using the following theories: LDA with no SO (dotted lines), LDA with SO effects taken into account 
(dashed lines) and $GW$ with SO interactions (solid lines).
(c,d) Degree of band inversion for the valence band (Eq.~\ref{BI}) of Bi$_2$Se$_3$ 
and Bi$_2$Te$_3$, respectively, calculated along the same $k$-path as in panels (a) and (b).  
}
\end{figure}

We start our discussion by considering the $GW$ quasiparticle energy corrections
$\Delta E_{\rm QP}(n,{\bf k})=E_{GW}(n,{\bf k})-E_{\rm LDA}(n,{\bf k})$ for bulk Bi$_2$Se$_3$ and Bi$_2$Te$_3$. Figs.~\ref{fig1}(a,b)
show $\Delta E_{\rm QP}(n,{\bf k})$ evaluated on a 6$\times$6$\times$6 $k$-point grid as a function of LDA energy
with no SO interactions taken into account. For convenience, we set the energies corresponding
to the valence band maximum (VBM) as a reference (i.e. $\Delta E_{\rm QP}({\rm VBM}) = E_{\rm LDA}({\rm VBM}) = 0$ for the case with 
no SO interactions included). When SO interactions are neglected both materials are direct 
band gap semiconductors with VBM and conduction band minimum (CBM) located at the $\Gamma$ 
point [dotted lines in Figs.~\ref{fig2}(a,b)]. The inclusion of $GW$ self-energy corrections
increases the LDA direct gaps at $\Gamma$ point (which are 0.151~eV and 0.188~eV for Bi$_2$Se$_3$ and Bi$_2$Te$_3$, respectively) 
by 0.212~eV and 0.277~eV, respectively. After SO interactions have been introduced in both LDA 
and $GW$ calculations, the values of $\Delta E_{\rm QP}(n,{\bf k})$ barely change except for those which correspond 
to VBM and CBM in Figs.~\ref{fig1}(a,b). These states, shown as large circles in Figure~\ref{fig1}, 
change their order in energy.


\begin{figure}
\includegraphics[width=7.5cm]{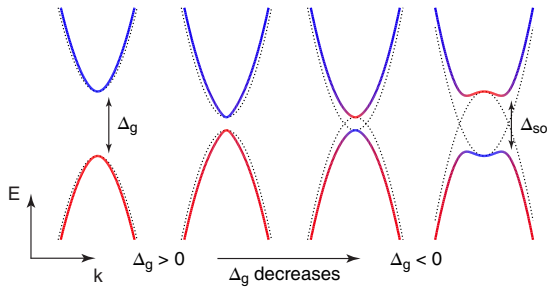}
\caption{\label{fig3}
(color online). Evolution of the band structure (solid lines) calculated for model 
Hamiltonian given by Eq. (\ref{matrix}) upon the decrease of parameter $\Delta_{\rm g}$ (see text). 
Dotted lines represent the solution calculated in the absence of off-diagonal (SO) matrix elements.  
}
\end{figure}

The observed seemingly counterintuitive behavior is a direct consequence of band inversion 
due to SO interactions. It can be illustrated using the $k \cdot p$ Hamiltonian for Bi$_2$Se$_3$-type materials \cite{Zhang09}:
\begin{equation}
\label{matrix}
{\mathcal H}_{k \cdot p}(k) = \left( \begin{array}{cccc}
{\mathcal M}({\bf k}) & A_{z} k_{z}               &               0 & A_{xy} k_{-} \\
A_{z} k_{z}               & -{\mathcal M}({\bf k}) & A_{xy} k_{-} & 0 \\
0               & A_{xy} k_{+} & {\mathcal M}({\bf k})& -A_{z} k_{z} \\
A_{xy} k_{+} & 0               & -A_{z} k_{z}               & -{\mathcal M}({\bf k}) 
\end{array} \right) 
+ \epsilon_0 (k)
\end{equation}
with $k_{\pm} = k_x \pm ik_y$ and $k=|{\bf k}|$. Without loss of generality we assume ${\mathcal M}({\bf k}) = -\Delta_{\rm g}/2-k^2/2m^*$ with
a single $m^*$ parametrizing both the valence and conduction bands, $\epsilon_0 (k) = 0$, and $A_{xy} = A_z = A$.
As the parameter $\Delta_{\rm g}$ decreases, a band inversion takes place around $k = 0$ for $\Delta_{\rm g} < 0$ and 
the band gap closes in the absence of off-diagonal matrix elements (dotted lines in Fig.~\ref{fig3}). 
However, these off-diagonal SO matrix elements ensure non-zero band gap even in the band-inversion regime. 
For a small magnitude $\Delta_{\rm g} < 0$ the bands remain parabolic and the gap is $-\Delta_{\rm g}$.
Further decrease of $\Delta_{\rm g}$ leads to the ``camelback'' shaped bands and the band gap is $\Delta_{\rm SO} \gtrsim 2 A k$.
The ``camelback'' feature is clearly seen for the valence bands in the LDA band structures of both Bi$_2$Se$_3$ and Bi$_2$Te$_3$
[dashed lines in Fig.~\ref{fig2}(a,b)]. The physical effect incorporated in the quasiparticle self-energy 
correction is the increase of $\Delta_{\rm g}$ 
which is typically underestimated in DFT. Upon an increase of the value $\Delta_{\rm g}$ in the $\Delta_{\rm g} < 0$ regime, the 
energy of the valence band at $k = 0$ increases while the energy of the conduction band at $k = 0$
decreases. This behavior is the opposite to the ``normal'' situation where no band inversion takes place.
In other words, the quasiparticle self-energy corrections to the inverted bands reduce the direct Kohn-Sham DFT gap at $\Gamma$.
This is exactly what is observed in Figs.~\ref{fig1}(c,d) (large symbols) and in Figs.~\ref{fig2}(a,b) at $\Gamma$.
 
The band structures of bulk Bi$_2$Se$_3$ and Bi$_2$Te$_3$ including both SO interactions and
the $GW$ quasiparticle self-energy corrections [solid lines in Fig.~\ref{fig2}(a,b)] have been 
calculated using the Wannier interpolations technique \cite{Marzari97,Mostofi08}. 
Interestingly, the LDA band gap of bulk Bi$_2$Se$_3$ (0.29~eV) barely changes after the inclusion of 
$GW$ corrections (0.30~eV). However, its character changes from indirect to direct in agreement with recent 
experiments \cite{Chen10}. The surprising accuracy of LDA in predicting the magnitude of minimum band gap is fortuitous.
In Bi$_2$Te$_3$, the LDA and $GW$ indirect band gaps are 0.09~eV and 0.17~eV \cite{Kioupakis10}, with the latter 
being in good agreement with experiment \cite{Chen09}. One noticeable effect of the $GW$ corrections on 
band dispersion is a considerable diminution of the dip in the valence bands at the $\Gamma$ 
point. This behavior is also consistent with the discussed two-band model [Eq.~(\ref{matrix}); Fig.~\ref{fig3}].

In order to gain further understanding of the effects of $GW$ corrections we define the degree of
band inversion for the valence band
\begin{equation}
\label{BI}
{\rm BI}_{\rm VB}({\bf k}) = \sum_{i = {\rm VB}; j \in {\rm unocc.}} a^*_{ij}({\bf k}) a_{ij}({\bf k}),
\end{equation}
where the eigenfunctions $\psi_i({\bf k}) = \sum_j a_{ij}({\bf k}) \phi_j({\bf k})$ of Hamiltonian 
${\mathcal H}({\bf k})={\mathcal H}_0 ({\bf k}) + {\mathcal H}_{\rm SO}({\bf k})$ are expressed 
in terms of $\phi_j({\bf k})$, the eigenfunctions of ${\mathcal H}_0({\bf k})$ which does not
include SO interactions. The results for both LDA and $GW$ methods are plotted in Figs.~\ref{fig2}(c,d).
For both Bi$_2$Se$_3$ and Bi$_2$Te$_3$ band inversion
takes place only in a limited region of the Brillouin zone around the $\Gamma$ point where ${\rm BI}_{\rm VB}$ 
achieves almost 100\%. The introduction of $GW$ shifts somewhat reduces the extension of this region
of band inversion, but at the $\Gamma$ point it is still complete. A very similar picture was obtained for 
the degree of band inversion of the conduction band (not shown).


The crux of our study is, of course, to investigate the effects of $GW$ quasiparticle self-energy 
corrections on the topological surface states. Addressing this problem in a straightforward way 
would require performing $GW$ calculations for two-dimensional slab models. At present, it is computationally too demanding 
to perform converged $GW$ calculations on systems of this size. To overcome this
difficulty, we parametrize the quasiparticle corrections $\Delta E_{\rm QP}$ in terms of energy-dependent 
``scissors operators'' $\Delta \tilde{E}_{\rm QP} = aE_{\rm LDA}^2 + bE_{\rm LDA} + c$ 
using the results of our $GW$ calculations for bulk materials with no SO interactions included.
Valence and conduction bands are fitted separately. Additionally, we require our ``scissors operators''
to reproduce exactly the quasiparticle shifts of the VBM and CBM, the critical components in our consideration.
The fitted functions are shown in Figs.~\ref{fig1}(a,b), and the corresponding parameters are given in 
Ref.~\onlinecite{scissors}. 


\begin{figure}[b]
\includegraphics[width=7.5cm]{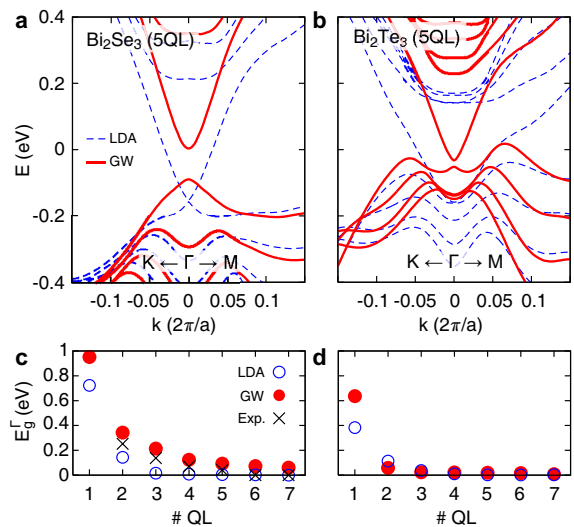}
\caption{\label{fig4}
(color online). Band structures calculated for 5QL slabs of (a) Bi$_2$Se$_3$ and (b) Bi$_2$Te$_3$
using LDA and $GW$ (SO interactions included in both cases). (c,d) Band gaps at the $\Gamma$ point 
as a function of slab thickness for Bi$_2$Se$_3$ and Bi$_2$Te$_3$, respectively. Experimental 
results are reproduced from Ref.~\onlinecite{Zhang10}. The zero of energy is set at $E_{\rm LDA}({\rm VBM})$ with
no SO.
}
\end{figure}

Figures~\ref{fig4}(a,b) show the band structures in the vicinity of the $\Gamma$ point computed for
(111) slabs of 5-quintuple-layers (5QL) thickness using plain LDA and after applying the proposed generalized
``scissors operator'' technique.
For the latter, we apply the ``scissors operators'' to the LDA Hamiltonian without SO interactions in 
the Bloch-state basis to obtain the quasiparticle Hamiltonian. The SO coupling terms are then added
to it and the Hamiltonian is diagonalized to obtain the final quasiparticle energies.
Both methods give rise to topologically protected surface states appearing as characteristic
``Dirac cone'' features at the $\Gamma$ point but with a gap $E_g^\Gamma$ owing to the hybridization 
of the surfaces states at the opposite surfaces of thin slabs \cite{Liu10,Yazyev10}.
The inclusion of quasiparticle corrections results in the following two important changes. First, the 
magnitudes of $E_g^\Gamma$ are enlarged, especially in Bi$_2$Se$_3$ [Figs.~\ref{fig4}(c,d)].
While LDA predicts essentially zero values of $E_g^\Gamma$ for slabs thicker than 3QL, the dependence
turns into a slow 1/width decay after the ``scissors operator'' was applied 
[Fig.~\ref{fig4}(c)]. This behavior is actually consistent with the gaps measured in thin
films of Bi$_2$Se$_3$ on SiC substrate \cite{Zhang10}. The experimental magnitudes are somewhat smaller
which can be attributed to enhanced screening due to the presence of the substrate which is neglected in our
calculations. The changes of band gaps are less systematic in the case of Bi$_2$Te$_3$. For slab thickness
larger than 4QL, the magnitudes of $E_g^\Gamma$ are larger in the calculations including the quasiparticle 
corrections. Second, the positions of Dirac points relative to the bulk bands change significantly. For both 
materials, the quasiparticle corrections ``lift'' the Dirac point from the bulk valence band. By extrapolating 
the results of our calculations on Bi$_2$Se$_3$ to the infinite slab thickness, we find that the incorporation of
quasiparticle corrections via the ``scissors operator'' technique changes the position of Dirac point
from 0.04~eV {\it below} the bulk valence band \cite{Yazyev10} to 0.07~eV {\it above} it. The latter value 
is in better agreement with experimental results of Analytis {\it et al.}: 0.205~eV below the bulk conduction 
band or 0.095~eV above the bulk valence band assuming a 0.30~eV band gap \cite{Analytis10}. 
For Bi$_2$Te$_3$, the quasiparticle corrections change the Dirac point energy from $-$0.20~eV to $-$0.10~eV
relative to the bulk VBM. The experimentally observed value is $-$0.13~eV \cite{Chen09}.  
On the contrary, we find that some properties are not affected by the quasiparticle corrections. For instance, the degree 
of spin polarization of surface states investigated in Ref.~\cite{Yazyev10} change very little.


In conclusion, while the Kohn-Sham DFT band structures are able to provide qualitative description of the topologically
nontrivial electronic structure of Bi$_2$Se$_3$ and Bi$_2$Te$_3$, a quantitative 
agreement with a number of experimentally measured properties is achieved only after 
including the $GW$ quasiparticle self-energy corrections. We further propose an energy-dependent ``scissors 
operator'' technique which allows the introduction of parametrized quasiparticle corrections 
into standard DFT calculations before SO interactions are included, thus greatly enhancing their predictive power 
in describing systems based on the discussed topological insulators.


This work was supported in part by NSF Grant No.~DMR10-1006184. O.\ V.\ Y. is recipient of 
a Swiss NSF fellowship (grant No.~PBELP2-123086) and partially supported by the Director, 
Office of Science, Office of Basic Energy Sciences, 
Division of Materials Sciences and Engineering Division, U.S. Department of Energy under 
Contract No.~DE-AC02-05CH11231. J.\ E.\ M. acknowledges support from the Center on Functional 
Engineered Nano Architectonics. The computation of the quasiparticle corrections was done 
with codes supported by NSF. Computational resources have been provided by TeraGrid (Kraken).


\begin{references}

\bibitem{reviews}
J. E. Moore, Nature (London) {\bf 464}, 194 (2010);
M. Z. Hasan and C. L. Kane, Rev. Mod. Phys. {\bf 82}, 3045 (2010);
X.-L. Qi and S.-C. Zhang, arXiv:1008.2026 (2010).

\bibitem{Xia09}
Y. Xia {\it et al.}, Nature Phys. {\bf 5}, 398 (2009).

\bibitem{Zhang09}
H. Zhang {\it et al.}, Nature Phys. {\bf 5}, 438 (2009).

\bibitem{TI_predictions}
H. Lin {\it et al.}, Phys. Rev. Lett. {\bf 105}, 036404 (2010);
B. Yan, {\it et al.}, Europhys. Lett. {\bf 90}, 37002 (2010);
S. V. Eremeev, Yu. M. Koroteev, and E. V. Chulkov, JETP Lett. {\bf 91}, 594 (2010);
S. Chadov {\it et al.}, Nature Mater. {\bf 9}, 541 (2010);
H. Lin {\it et al.}, Nature Mater. {\bf 9}, 546 (2010);
Y. Sun, X.-Q. Chen, S. Yunoki, D. Li, Y. Li, Phys. Rev. Lett. {\bf 105}, 216406 (2010);
W. Zhang {\it et al.}, Phys. Rev. Lett. {\bf 106}, 156808 (2011);
F. Virot {\it et al.}, Phys. Rev. Lett. {\bf 106}, 236806 (2011),
among many others.

\bibitem{BiTlX2}
T. Sato et al., Phys. Rev. Lett. 105, 136802 (2010);
K. Kuroda et al., Phys. Rev. Lett. 105, 146801 (2010).

\bibitem{Louie06}
S. G. Louie, in {\it Conceptual Foundations of Materials: A Standard
Model for Ground- and Excited-State Properties}, edited by
S. G. Louie and M. L. Cohen (Elsevier, Amsterdam, 2006), p. 9.

\bibitem{Vidal11}
J. Vidal, X. Zhang, L. Yu, J.-W. Luo, and A. Zunger,
Phys. Rev. B {\bf 84}, 041109 (2011).

\bibitem{Hedin70}
L. Hedin and S. Lundqvist, 
Solid State Physics {\bf 23}, 1 (1970).

\bibitem{Hybertsen86a}
M. S. Hybertsen and S. G. Louie, 
Phys. Rev. B {\bf 34}, 5390 (1986).

\bibitem{QE}
P. Giannozzi {\it et al.},
J. Phys.: Condens. Matter {\bf 21}, 395502 (2009).

\bibitem{Troullier91}
N. Troullier and J. L. Martins, 
Phys. Rev. B {\bf 43}, 1993 (1991).

\bibitem{BerkeleyGW}
J.\ Deslippe {\it et al.}, {\sc BerkeleyGW} package. For additional information see http://www.berkeleygw.org.

\bibitem{Hybertsen86b}
M. S. Hybertsen and S. G. Louie, 
Phys. Rev. B {\bf 34}, 2920 (1986).

\bibitem{Yazyev11}
O. V. Yazyev, G. Samsonidze, and S. G. Louie, unpublished (2011). Note, 
we do not expect this agreement to hold for all TIs as band inversion 
may in some cases have significant effects on the dielectric matrix. 

\bibitem{Marzari97}
N. Marzari and D. Vanderbilt, 
Phys. Rev. B {\bf 56}, 12847 (1997).

\bibitem{Mostofi08}
A. A. Mostofi, J. R. Yates, Y.-S. Lee, I. Souza, D. Vanderbilt, and N. Marzari, 
Comput. Phys. Commun. {\bf 178}, 685 (2008).

\bibitem{Chen10}
Y. L. Chen {\it et al.}, 
Science {\bf 329}, 659 (2010).

\bibitem{Kioupakis10}
E. Kioupakis, M. L. Tiago, and S. G. Louie,
Phys. Rev. B {\bf 82}, 245203 (2010).

\bibitem{Chen09}
Y. L. Chen {\it et al.}, 
Science {\bf 325}, 178 (2009).

\bibitem{scissors}
For Bi$_2$Se$_3$, valence bands: $a$ = 0.0128~eV$^{-1}$, $b$ = 0.112, $c$ = 0~eV;
               conduction bands: $a$ = 0.0037~eV$^{-1}$, $b$ = 0.011, $c$ = 0.233~eV.
For Bi$_2$Te$_3$, valence bands: $a$ = 0.0079~eV$^{-1}$, $b$ = 0.067, $c$ = 0~eV;
               conduction bands: $a$ = 0.0039~eV$^{-1}$, $b$ = $-$0.009, $c$ = 0.314~eV.
We recall that $\Delta \tilde{E}_{\rm QP}({\rm VBM}) = \Delta E_{\rm QP}({\rm VBM})= E_{\rm LDA}({\rm VBM}) = 0$ 
and $\Delta \tilde{E}_{\rm QP}({\rm CBM}) = \Delta E_{\rm QP}({\rm CBM})$ by convention. 
Constant energy shifts $c$ are the leading corrections.

\bibitem{Liu10}
C.-X. Liu {\it et al.},
Phys. Rev. B {\bf 81}, 041307 (2010).

\bibitem{Yazyev10}
O. V. Yazyev, J. E. Moore, and S. G. Louie,
Phys. Rev. Lett. {\bf 105}, 266806 (2010).

\bibitem{Zhang10}
Y. Zhang {\it et al.}, Nature Phys. {\bf 6}, 584 (2010).

\bibitem{Analytis10}
J. G. Analytis {\it et al.}, 
Phys. Rev. B {\bf 81}, 205407 (2010).

\end{references}
\end{document}